\documentclass[aps,apl]{revtex4}
\usepackage{amsmath,amssymb}
\input epsf
\usepackage{verbatim}
\usepackage{graphicx}
\usepackage{bm}
\usepackage{natbib}
\def\pmb#1{\setbox0=\hbox{#1}
\kern-.025em\copy0\kern-\wd0 \kern-.05em\copy0\kern-\wd0
\kern-.025em\raise.0433em\box0}

\newcommand{\C}{\mathbb C}

\newcommand{\beq}{\begin{equation}}
\newcommand{\eeq}{\end{equation}}
\newcommand{\ba}{\begin{eqnarray}}
\newcommand{\ea}{\end{eqnarray}}

\input epsf

\usepackage[english]{babel}
\begin{document}

\title[]{Control of Rayleigh-like waves in thick plate Willis metamaterials} 


\author{Andr\'e Diatta$^{1}$,Younes Achaoui$^{1}$,  St\'ephane Br\^ul\'e$^{2}$, Stefan Enoch$^{1}$ and S\'ebastien Guenneau$^{1}$}
\affiliation{1. Aix$-$Marseille Univ., CNRS, Centrale Marseille, Institut Fresnel, 13013 Marseille, France\\  
2. Dynamic soil laboratory, M\'enard, 91620 Nozay, France
}

\begin{abstract}
Recent advances in control of anthropic seismic sources in structured soil led us to explore interactions of elastic waves
propagating in plates (with soil parameters)  structured with concrete pillars buried in the soil. Pillars are ${2}$ {m in diameter}, ${30}$ m in depth and
the plate is ${50}$ m in thickness. {We study} the frequency range ${5}$ to ${10}$ Hz, {for which} Rayleigh wave wavelengths are smaller than the plate thickness. This frequency range is compatible with frequency ranges of particular interest in earthquake engineering. It is demonstrated in this paper that two seismic cloaks' configurations allow for an unprecedented flow of elastodynamic energy associated with Rayleigh surface waves. {The first cloak design is inspired by some approximation of ideal cloaks' parameters within the framework of thin plate theory. The second, more accomplished but more involved, cloak design is deduced from a geometric transform in the full Navier equations that preserves the symmetry of the elasticity tensor but leads to Willis' equations, well approximated by a homogenization procedure, as corroborated by numerical simulations. The two cloaks's designs are strikingly different, and the superior efficiency of the second type of cloak emphasizes the necessity for rigor in transposition of existing cloaks's designs in thin plates to the geophysics setting.} Importantly, we focus our attention on geometric transforms applied to thick plates, which is an intermediate case between thin plates and semi-infinite media, not studied previously. Cloaking efficiency (reduction of the disturbance of the wave wavefront and its amplitude behind an obstacle) and protection (reduction of the wave amplitude within the center of the cloak) are studied for ideal and approximated cloaks' parameters. These results represent a {preliminary} step towards designs of seismic cloaks for surface Rayleigh waves propagating in sedimentary soils structured with concrete pillars.
\end{abstract}
\maketitle

\section{Introduction}
In a recent theoretical proposal \cite{diatta2014}, control of body waves was numerically demonstrated with a spherical shell consisting of an anisotropic heterogeneous elasticity tensor without the minor symmetries (what falls within the framework of so-called Cosserat media), and an heterogeneous (but isotropic) density. In \cite{diatta2014}, it was suggested that a simplified version of this elastodynamic cloak could be implemented to protect objects, nuclear waste or even large scale infrastructures buried deep down in the soil. However, achieving asymmetric elasticity tensors for instance via effective medium approaches remains a challenge. Indeed, the Cosserat theory \cite{cosserat} of elasticity (or micropolar elasticity) incorporates a local rotation of points as well as the translation assumed in classical elasticity, and a couple stress (a torque per unit area) as well as the force stress (force per unit area). In the isotropic Cosserat solid or micropolar continuum, there are six elastic constants, in contrast to the classical elastic solid which is described by two constants. This makes Cosserat cloaks fairly challenging to engineer. In the present paper, we follow a different route towards seismic wave cloaking for surface Rayleigh waves, without resorting to Cosserat media.
{Before we move to the core material of this paper}, we would like to first recall why soils structured at a meter scale might counteract deleterious action of certain types of seismic waves, what might seem at first glance fairly counter-intuitive. 

More than a million earthquakes are recorded every year, by a worldwide system of earthquake detection stations, some of which are particularly devastating and cause human casualties, such as the earthquake of magnitude-6.2 that struck Italy on the 22nd of August 2016 at 01:36 GMT, $100$ km north-east of Rome, not far from L'Aquila, where a similar earthquake struck on the 9th of April 2009. The propagation velocity of the seismic waves depends on density and elasticity of the earth materials (clearly, it is much different in sedimentary soils and rocks). At the scale of an alluvial basin, such as in L'Aquila, seismic effects involve various phenomena, such as wave trapping, resonance of whole basin, propagation in heterogeneous media, and the generation of surface waves at the basin edge \cite{brule}. As noted in \cite{eml2016}, wherein meter-scale inertial resonators were introduced to reflect surface and body seismic waves, due to the surface wave velocity in superficial and under-consolidated recent material (less than $100$ to $300$ m.s$^{-1})$, wavelengths of surface waves induced by natural seismic sources or construction work activities are shorter than those of earthquake generated direct P (primary, i.e. longitudinal compressional) and S (secondary, i.e. transverse shear) waves (considering the $0.1$-$50$ Hz frequency range), from a few meters to a few hundreds of meters.
These are of similar length to that of buildings, therefore leading to potential building resonance phenomena in the case of earthquakes such as in L'Aquila. Other sources of
soil vibrations one may wish to attenuate include traffic and construction works \cite{semblat}, and screening methods have been proposed to achieve that \cite{woods,banerjee}.

Interestingly, back in 1999 Rayleigh wave attenuation was theoretically and experimentally achieved in marble quarry with air holes displaying kHz stop bands \cite{meseguer} and similar filtering effects in a microstructured piezoelectric for MHz till $1$ GHz surface waves \cite{sarah,younes1,younes2,younes3,younes4}. In our previous work on control of surface Rayleigh waves in soils structured with borehole inclusions, we used so-called elastic stop bands which are frequency intervals for which waves are disallowed to propagate in an infinite periodic array through Bragg interferences (when the wave wavelength is on the order of the array pitch). With boreholes $0.3$m in diameter, with a center to center spacing of $1.7$m and the soil parameters on this field experiment \cite{prl2014}, this meant we had a stop partial band around $50$ Hz (for only one crystallographic direction). Three rows of boreholes were enough to reduce the energy of the signal by about $30$ per cent behind the seismic metamaterial. However, if one were to design a shield for the frequency interval $5$ to $10$ Hz, it would be necessary to scale up the array of boreholes by a factor of five to ten, which requires a lot of free space around the area, one wishes to protect. To overcome this obstacle, we now propose to use columns of concrete instead of boreholes, so as to achieve subwavelength control of Rayleigh surface waves. We show in figure \ref{Menard} an example of such a structured soil, which was designed by civil engineers of the Menard company.

Importantly, the contrast between the soil and concrete column's parameters makes it possible to achieve uncommon effective material parameters, such as with strong artificial anisotropy and dispersion, and even artificial inertia and viscosity that can be interpreted as rank-3 tensors in an effective Willis equation. This leads to a markedly enhanced control of surface Rayleigh wave trajectories compared to our earlier work \cite{prl2014}. Indeed, comparisons between numerical results demonstrate that one can achieve some cloaking of surface (Rayleigh-like) wave trajectories, where the wave field is almost unperturbed outside the seismic cloak, whereas it {simultaneously} nearly vanishes in its center. Bearing in mind that we use soil and concrete elastic parameters for the thick plate and pillars, and that we consider wave frequencies lower than $10$ hertz, we claim that our results represent a preliminary step towards implementation of seismic cloaks in civil engineering.   

In this paper, we would like to present a novel approach for the design of seismic cloaks, which is based on the Willis-type equations, as proposed in the seminal paper by Milton, Briane and Willis \cite{milton2006}, which investigated form invariant governing equations in physics, notably coordinates transforms in Navier equations ensuring a transformed symmetric elasticity tensor. For the sake of simplicity in the numerical computations and physical discussions, we shall restrict ourselves to elastic waves in thick plates, rather than in semi-infinite media. Such a choice is in compliance with the fact that Rayleigh wave amplitudes decay exponentially {(it vanishes within two wavelengths)} with depth, becoming approximatively zero by the time a depth of two wavelengths has been reached.
 However, we consider wavelengths small compared to the plate's thickness, hence surface waves are akin to Rayleigh, rather than Lamb, waves. Indeed, as noted in \cite{prl2014}, the elastic plate model already gives some interesting insight in the physics of seismic waves. 
The frequency of interest in earthquake engineering is governed by the frequencies of structures for the fundamental mode and the first harmonics. Typically, the fundamental period expressed in seconds, for N-story structure could be approximated by $N/10$. For our numerical implementations, we have decided to focus on the range of $5$ to $10$ Hz corresponding to low-rise common building with only few storeys.

The plan of the paper is as follows: We first present some derivation of transformed Navier equations with a symmetric elasticity tensor, for thick plates. The obtained Willis equations are then simplified (removal of rank-3 tensors) and the illustrative case of a cloak with heterogeneous anisotropic Willis equations is numerically solved with finite elements implemented in the COMSOL commercial package. We then explain how one can approximate these anisotropic elastic parameters with an effective medium approach. The touchstone of our homogenization approach is that all tensors are symmetric.
 We finally propose two designs of soils structured with concrete pillars which are numerically validated with finite elements. 
\section{Elastodynamic Willis Material, equations of motion}
The propagation of
elastic waves is governed by the Navier equations. Assuming
time harmonic $\exp(-i\omega t)$ dependence, with $\omega$ as the
angular wave frequency and $t$ the time variable, allows us to work directly in the spectral domain.
Such dependence is assumed henceforth and suppressed, leading to
\begin{equation}
\nabla\cdot{\boldsymbol{\sigma}}=-i\omega {\bf p} \; , \; \; \;  \;  \;
\boldsymbol{\sigma}={\C}:\nabla{\bf u} \; ,  \; \; \; \;  \; 
{\bf p}=-i\omega\rho {\bf u} \; , \;
\label{navier}
\end{equation}

where $\rho$ is the density of the (possibly heterogeneous
isotropic) elastic medium and ${\bf u}=(u_r,u_\theta,u_z)$
is the three-component displacement field in a cylindrical
coordinate basis ${\bf x}=(r,\theta,z)$. Also,  $\C$ is the rank-four
(symmetric) elasticity tensor with components
$C_{ijkl}$ ($i,j,k,l=r,\theta,z$) and $:$ stands for the double contraction between tensors.  For example, ${\C}:\nabla{\bf u}$ is the 2-tensor with components  $({\C}:\nabla{\bf u})_{ij}=C_{ijkl} \Big(\nabla{\bf u}\Big)_{kl}$.
Let us consider a coordinate change ${\bf x}\longmapsto {\bf x}'$,
where ${\bf x'}=(r',\theta',z')$ are stretched spherical coordinates.
In general, this  leads to
a transformed equation \cite{milton2006,norris2011}

\begin{equation}
\begin{array}{ll}
\nabla'\cdot\Big({{\C}':\nabla'{\bf u'}+{\bf S}\cdot{\bf u'}}\Big)=-i\omega \Big({{\bf D}:\nabla'{\bf u'}-i\omega{\boldsymbol\rho}{\bf u}'}\Big) \; , \;  \; \; \;  \;
{\bf u'}={\bf A}^{-T}{\bf u} \; ,
\end{array} 
\label{navierwillis}
\end{equation}
or, in a more compact way
 \begin{equation}
\begin{array}{ll}
\nabla'\cdot{\boldsymbol{\sigma}'}=-i\omega {\bf p}' \; , \;  \; \; \;  \;
\boldsymbol{\sigma}'={\C}':\nabla'{\bf u'}+{\bf S}\cdot{\bf u'} \; , \; \\
{\bf p}'={\bf D}:\nabla'{\bf u'}-i\omega{\boldsymbol\rho}{\bf u}' \; , \;  \; \; \;  \; 
{\bf u'}={\bf A}^{-T}{\bf u} \; ,
\end{array} 
\label{navierwillis2}
\end{equation}
where ${\bf S}$, ${\bf D}$  are 3-tensor fields, $\nabla'$ is the gradient in transformed coordinates ${\bf x'}$ and
${\bf u}'({\bf x'})={\bf u'}(r',\theta',z')$ is a transformed displacement
in stretched cylindrical coordinates.
Note that the transformed stress is generally not symmetric.
Note also that in general ${\bf A}$ is a matrix field.
In order to preserve
the symmetry of the stress tensor, one
assumes that  ${\bf A}$ is a multiple ${\bf A}=\xi\partial{\bf x}'/\partial{\bf x}$, of the Jacobian matrix $\partial{\bf x}'/\partial{\bf x}$ of the transformation, 
where $\xi$ is a non-zero scalar, in which case (\ref{navierwillis}),  (\ref{navierwillis2}) are said to be  Willis-type equations \cite{willis1981,milton2006}.

\section{Approximated cloaking with Willis material without 3-tensors}



Following the proposal of Pendry et al. of an invisibility cloak via transformation optics \cite{pendry}, we consider a specific geometric transformation, in cylindrical coordinates
\begin{eqnarray}
r'= 
\left\{
\begin{array}{lr}
r_1+\frac{r_2-r_1}{r_2}r , \label{PTransform2}   
    \text{ if  } 0\leq r\leq r_2
 \\
r   \text{ if  } r \ge r_2
\end{array}
\right.
\end{eqnarray}
with
$
 \theta'=\theta, z'=z
$.
Such a transformation blows the axis $\{(r,\theta, z), ~~ r=0 \}$ of a vertical solid cylinder of radius $r_2$, into the cylinder  $\{(r,\theta, z), ~~ r=r_1 \},$ while confining the whole solid cylinder  $\{(r,\theta, z), ~~ r\le r_2 \}$ into the following  hollow cylinder  $\{(r,\theta, z), ~~ r_1\le r\le r_2 \}$ of inner and outer radii $r_1$ and $r_2$ respectively.   

In the sequel, we plug the  transformation (\ref{PTransform2}) into the Navier equations 
\begin{eqnarray}\label{navier1}
\nabla\cdot{\C}:\nabla{\bf u}+\rho\omega^2 {\bf u}={\bf 0},
\label{snavier}
\end{eqnarray}
to  derive the corresponding transformed  equations (\ref{navierwillis}), (\ref{navierwillis2}).
 From  here on, we drop (neglect) the 3-order tensors ${\bf S}$  and {\bf D}  and hence use an approximate version of the Willis-like Equations. 
 Note that in this case, 
${\C}'$ obviously still has its minor and major symmetries, but is anisotropic and heterogeneous, with coefficients in cylindrical coordinates


\begin{eqnarray}
C_{r'r'r'r'}'&=&\Big(\frac{r_2-r_1}{r_2}\Big)^2~\frac{r'-r_1}{r'}~(\lambda+2\mu); ~~
   C_{r'r'\theta\theta}'=   C_{\theta\theta rr}'= \Big(\frac{r_2-r_1}{r_2}\Big)^2\frac{r'}{r'-r_1}\lambda;
\nonumber
\\
  C_{r\theta r\theta}'&=&C_{r'\theta\theta r}'=  C_{\theta rr\theta}'= C_{\theta r\theta}'= \Big(\frac{r_2-r_1}{r_2}\Big)^2\frac{r'}{r'-r_1}\mu; 
 \nonumber
\\
  C_{rrzz}'&=&C_{zzrr}'= \frac{r'-r_1} {r'} \lambda; 
 \nonumber
\\
C_{rzrz}'&= & C_{rzzr}'=C_{zrrz}'=C_{zrzr}'= \frac{r'-r_1}{r'} \mu ;  \nonumber
\\
   C_{\theta\theta\theta\theta}'&=& \Big(\frac{r_2-r_1}{r_2}\Big)^2~\Big(\frac{r'}{r'-r_1}\Big)^3~(\lambda+2\mu); ~~  C_{\theta\theta zz}'=  C_{zz\theta\theta}'  = \frac{r'} {r'-r_1} \lambda ;
  \nonumber
\\
  C_{\theta z\theta z}'&=& C_{\theta zz\theta}'=C_{z\theta\theta z}'=C_{z\theta z\theta}'=\frac{r'} {r'-r_1} \mu;~~
  \nonumber
\\
 C_{zzzz}'&=&\Big(\frac{r_2}{r_2-r_1}\Big)^2\frac{r'-r_1}{r'}(\lambda+2\mu).
\label{ddcloaktensor} 
\end{eqnarray}
The stretched density
$\rho'$ is now a tensor field of order 2, in particular, it is anisotropic and inhomogeneous.  Namely, it has non-constant  different eigenvalues and is diagonal in cylindrical coordinates with components as follows

\begin{eqnarray}
\rho'&=& \text{diag} \Big(\rho'_{r'r'},~~ \rho'_{\theta\theta},~~\rho'_{zz}\Big)=\rho~\text{diag} \Big(\frac{r'-r_1}{r'},~~ \frac{r'}{r'-r_1},~~\frac{r_2^2}{(r_2-r_1)^2} \frac{r'-r_1}{r'}\Big).
\label{ddcloakdensor}
\end{eqnarray}

Although all the (minor and major) symmetries of $\C$ are preserved, the anisotropy in the azimuthal direction is infinite at the inner boundary of the cloak, as    
$C'_{\theta\theta\theta\theta}/C'_{r'r'r'r'}=\Big(\frac{r'}{r'-r_1}\Big)^4$ recedes to infinity as quickly as $\Big(\frac{1}{r'-r_1}\Big)^4$ when $r'$
  approaches $r_1.$  However $C'_{r'r'r'r'}$  and $ C_{zzzz}'$ are of the same order and both tend to zero at the same rate at the inner boundary $r'=r_1.$
Meanwhile, at the same boundary, the off-diagonal components 
$   C_{r'r'\theta\theta}',$   $C_{\theta\theta r'r'}',$ $ C_{r\theta r\theta}',$ $C_{r'\theta\theta r}',$ $ C_{\theta r'r'\theta}',$ $ C_{\theta r'\theta}',$  
$   C_{\theta\theta\theta\theta}',$ $ C_{\theta\theta zz}',$ $ C_{zz\theta\theta}', $ $ C_{\theta z\theta z}',$ $C_{\theta zz\theta}',$ $C_{z\theta\theta z}',$$C_{z\theta z\theta}',$  all become infinite at the rate $\frac{1}{r'-r_1}$, whereas 
 $ C_{r'r'zz}',$ $C_{zzr'r'}',$  $C_{r'zr'z}',$ $C_{r'zzr'}',$ $C_{zr'r'z}',$ $C_{zr'zr'}',$ $C_{\theta z\theta z}',$ $ C_{\theta zz\theta}',$ $C_{z\theta\theta z}',$ $C_{z\theta z\theta}',$  tend to zero. One remarks in the meantime that, the eigenvalues $\rho'_{r'r'}$ and $\rho'_{z'z'}$ of the density in the $r$ and $z$ directions both vanish, whereas in the  azimuthal direction $ \rho'_{\theta'\theta'}$ becomes infinite  at the inner boundary of the cloak. The material requirement for vanishing density along the vertical direction $\rho'_{z'z'}$  can be relaxed for Rayleigh waves, since their amplitude goes to zero within two wavelengths (this is fortunate as it means we won't have to structure the soil in the vertical direction when we approximate the ideal cloak parameters through homogenization approach). However, variation of transverse anisotropic density (in the horizontal xy-plane) according to (\ref{ddcloakdensor}) is a strong requirement for a fully operational cloak.


\subsection{Rayleigh-like waves in thick plates}
In this paper, we apply the approximate Willis cloaking discussed above,  to Rayleigh waves in particular in the framework of seismic metamaterials. 
Rayleigh waves, unlike Love waves, are surface waves that do not need a layer to guide them. The former surface waves exist in semi-infinite homogeneous isotropic media with a stress -free boundary. In our present work, we consider plates whose thickness is {larger} than the wavelength of the surface wave, which are akin to Rayleigh waves. When we make the geometric transform in the Navier equations, the stress-free boundary conditions also undergo the mapping. Hence one has  to apply the transformation to
 the Navier Equations coupled with the stress-free boundary condition
\begin{eqnarray}
\left\{
\begin{array}{lr}
\frac{\partial}{{\partial x_i}}\left({C}_{ijkl}\frac{\partial}{{\partial x_k}}{u}_l\right)+{\rho}_{jl}\omega^2 {u}_{l}=0 \;
,
\\
 {\bf n}\cdot (\C:\nabla {\bf u})=0~~ \text{on the boundary } ~~ z=z_0,
\end{array}
\right.
\end{eqnarray}
with ${\bf n}$  the (outward) normal at the points on the surface $z=z_0$.
In our case, the horizontal plane $z=z_0$ stands for the soil-air interface of the thick plate.
In the resulting Willis-type equations discussed in the previous sections, the approximated technique is again applied coupled with 

\begin{eqnarray}
 {\bf n'}\cdot (\C':\nabla' {\bf u})=0~~ \text{on the transformed boundary } ~~ z=z_0,
\end{eqnarray} 
where ${\bf n'}$ is the normal at the transformed boundary.

\subsection{Numerical implementation of Rayleigh-like waves in thick plates with soil parameters using Finite Element Method}

Aiming at applications particularly in civil engineering  such as seismic protection, we base the numerical model corresponding to the cloak described above on soil parameters.  Namely,  we use the Lam\'e coefficients  $\lambda=87.7*10^6$ Pa, $\mu=58.5*10^6$ Pa and density $\rho=1800$ kg/m$^3$. 
Although the approach  developed here is general, we chose to concentrate our numerical tests on the range of $5$ to $10$ Hz corresponding to low-rise common building with only few storeys.

The cloak consists of a vertical cylindrical hollow region of height $50$ m, with an annulus-like basis of inner and outer radii $r_1=10$ m  and $r_2=21.6$ m respectively and the $z-$axis as its main axis.  It surrounds a homogeneous isotropic cylindrical domain of radius $r_1$ (the cloaked region) made of soil.  Likewise, the  cloak itself is located inside a cylindrical homogeneous isotropic medium (the ambient space) also made of soil and  surrounded by a cylindrical shell of inner and outer radii $r_3=50$ m and $r_4=66.6$ m respectively, which is filled with an anisotropic heterogeneous absorptive medium acting as a (reflectionless) perfectly matched layer (PML). {The top plane} is a  stress-free surface, which is the  interface between the soil and the air.

\begin{figure}[h!]
\resizebox{84mm}{!}{\includegraphics{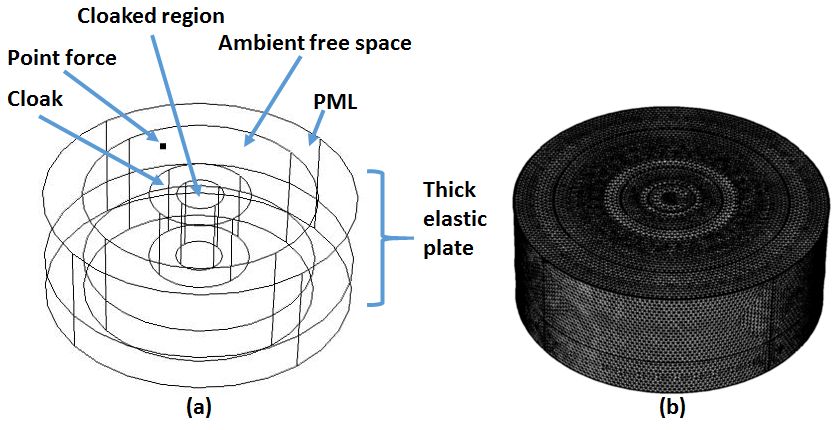}}
\vspace{-3mm}
\caption{(a) Geometric construction of the computational domain with a cylindrical Willis-like cloak surrounding a cylindrical region and a cylindrical annulus of PML to minimize wave reflections on the vertical boundary of the domain (the top  and bottom boundaries have stress-free data). (b) Mesh of the computational domain with $1,462,046$ tetrahedral elements, $54,726$ triangular elements, $1,660$ edge elements and  $41$ vertex elements.}\label{geometryaip}
\end{figure}

The  implementation of this model in the finite element package COMSOL MULTIPHYSICS requires the use of Cartesian coordinates. Fortunately, $40$ of the $3^4$ spatially varying entries of the transformed elasticity tensor  vanish identically. We mesh the computational domain using $1,462,046$ tetrahedral elements, $54,726$ triangular elements, $1,660$ edge
elements and  $41$ vertex elements, as shown in Fig. \ref{geometryaip}(b).

\begin{figure}[h!]
\resizebox{84mm}{!}{\includegraphics{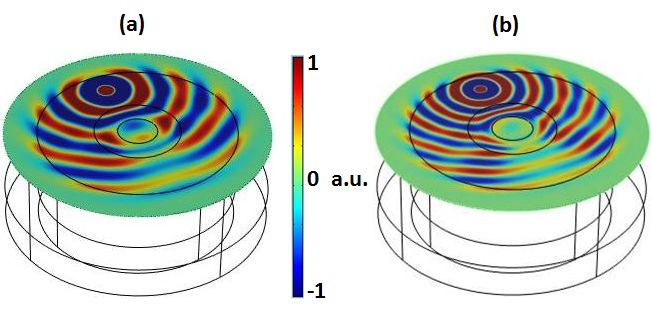}} 
\vspace{-3mm}
\caption{Out of plane component $u_z$ of the displacement field ${\bf u}=(u_r,u_\theta,u_z)$ generated by a point force vibrating at frequency $f=\omega/(2\pi)=5.5$ Hz in  panel (a) and $f=7.9$ Hz in (b),  polarized along $z$, which is located at the air-soil interface and at a distance $r=22.56$ m, from the axis of an approximated Willis cloak of inner radius $r_1=10$ m and outer radius $r_2=21.6$ m. One notes that the displacement field nearly vanishes in the invisibility region at the center of the cloak in panel B, whereas the amplitude and phase of the wave are nearly unperturbed outside the cloak. Such a seismic cloak for Rayleigh waves would protect a building placed in its center and have no impact on the surrounding buildings, unlike for a seismic shield like in \cite{prl2014} which would have disastrous effect for buildings facing reflected waves.
However, as already noted in \cite{diatta2016}, at certain frequencies corresponding to a countable spectrum of the stress-free cavity, trapped modes exist that would have an antagonistic effect, see panel (a).
This is reminiscent of trapped modes unveiled by the group of Greenleaf \cite{Greenleaf-prl-2008} in the context of quantum cloaks.}\label{rayleigh1aip}
\end{figure}

 A point force located at $(8.3 \hbox{m},36.7\hbox{m}, 16.7\hbox{m})$, - that is, it lies on the soil-air free surface at a distance of $37.6$ meters from the axis of the cloaked region, - oriented along the direction $(0,0,1)$ generates a Rayleigh-like wave at frequency  $5.5$ Hz (Fig \ref{rayleigh1aip} (A))-- which corresponds to a wave wavelength $\sim 33$ m that is greater than half of the basin depth ($25$ m), so this surface wave still behaves like a plate wave-- and at frequency $7.9$  Hz (Fig \ref{rayleigh1aip} (B))-- which corresponds to a wave wavelength $\sim 23$ m strictly lower than half of the basin depth-- so it does not feel the bedrock. Following predictions in \cite{diatta2016}, one expects that some Rayleigh-like waves generated by specific frequencies of the source will create resonances (trapped modes) within the cloaked region as in Figure \ref{rayleigh1aip}  (a), wherein a dipole mode can be observed (the inner cylinder wobbles) : This occurs at a countable set of eigenfrequencies corresponding to the problem of the Neumann (stress-free) cylindrical cavity in the center of the cloak, as we will investigate in section VI. At a source frequency away from the cavity trapped modes, one can clearly see that we achieve a good seismic protection (Fig. \ref{rayleigh1aip} (b)), see also Fig. \ref{figlast} where a quantitative study is carried out in the frequency range $3$ Hz to $10$ Hz. Elastic parameters for a cloak with a Willis-like heterogeneous anisotropic transformed medium seem unachievable in practice with current civil engineering techniques. We therefore present in the sequel a homogenization route towards achievable Willis-like seismic cloaks.
 
At this stage, let us point out that an alternative route to control of Rayleigh waves consists in applying tools of conformal optics as proposed by Ulf Leonhardt \cite{leonhardt},
that in the context of transformational elastodynamics lead to spatially varying isotropic elastic parameters \cite{andrea}. Although this might seem an easier way towards cloaking,
this requires structuring soil with pillars softer than soil, which is actually the opposite case to what we do in the sequel.
 
 \section{Homogenization approach for approximate Willis medium without rank-3 and rank-2 tensors}
Our homogenization approach to achieve required rank-4 and rank-2 tensors proceeds in two steps.
Firstly, we follow the proposal of Greenleaf and coworkers \cite{greenleaf2008} to apply the
two-scale convergence method of G. Allaire and G. Nguetseng \cite{2scale1,2scale2}  in the design of approximate multi-layered
cloaks, we derive a set of effective parameters for a multi-layered elastic cloak, which is akin to homogenized parameters obtained
by the G-convergence approach in \cite{jikov}, except that these parameters remain dependent upon the macroscopic variable. We consider
an alternation of concentric layers of respective elasticity tensors $\C_1$ and $\C_2$ and of respective densities $\rho_1$
and $\rho_2$ so that the overall elasticity tensor and density can be written
\begin{equation}
\begin{array}{ll}
\C_\eta({\bf x})=\C_1({\bf x})1_{[0,1/2]}(\frac{r}{\eta})+\C_2({\bf x})1_{[1/2,1]}(\frac{r}{\eta}) \\
\rho_\eta({\bf x})=\rho_1({\bf x})1_{[0,1/2]}(\frac{r}{\eta})+\rho_2({\bf x})1_{[1/2,1]}(\frac{r}{\eta})
\end{array}
\end{equation} 
where ${\bf x}$ is the position vector, and $r=\Vert{\bf x}\Vert=\sqrt{x_1^2+x_2^2}$ is its Euclidean norm and
$1_I$ is the indicator function of the interval $I$ ($I$ is typically the unit cell size along the direction of periodicity).
The spatially varying tensor $\C_\eta$ is periodic on $I=[0,1]$ and $\eta$ defines the periodicity (the thinner the layers,
the smaller $\eta$, the larger the number of layers). The governing equation reads:
\begin{equation}
\begin{array}{ll}
\rm{div}(\C_\eta({\bf x}) :\nabla {\bf u}_\eta({\bf x})) + \rho_\eta({\bf x})\omega^2 {\bf u}_\eta({\bf x}) ={\bf 0}
\end{array}
\end{equation} 
When $\eta$ tends to zero, it is shown in \cite{jikov} that the solution ${\bf u}_\eta$ to the above Navier equation two-scale converges towards ${\bf u}_0$ solution to the
homogenized Navier equation:
\begin{equation}
\begin{array}{ll}
\rm{div}(\C_0({\bf x}) :\nabla {\bf u}_0({\bf x})) + \rho_0({\bf x})\omega^2 {\bf u}_0({\bf x}) ={\bf 0}
\end{array}
\end{equation}
with an effective (symmetric) elasticity tensor
\begin{equation}
\begin{array}{ll}
2\xi\cdot\C_0({\bf x})\xi
&=\xi_{11}^2\left( 2<\mu>({\bf x}) - 4 <(\mu-\bar\mu)^2/(2\mu+\lambda)>({\bf x})\right)-4\xi_{11}tr\xi<(\mu-\bar\mu)(\lambda-\bar\lambda)/(2\mu+\lambda)>({\bf x}) \nonumber \\
&+{(tr\xi)}^2\left( <\lambda>({\bf x}) - <(\lambda-\bar\lambda)^2/(2\mu+\lambda)>({\bf x})\right)+4{<\mu^{-1}>}^{-1}({\bf x})\xi_{12}^2+2<\mu>({\bf x})\xi_{22}^2
\end{array}
\label{toto1}
\end{equation}
where
\begin{equation}
\bar\mu=\left<\frac{\mu}{2\mu+\lambda}\right>({\bf x}){\left<\frac{1}{2\mu+\lambda}\right>}^{-1}({\bf x}) \; , \; \bar\lambda=\left<\frac{\lambda}{2\mu+\lambda}\right>({\bf x})
{\left<\frac{1}{2\mu+\lambda}\right>}^{-1}({\bf x})
\end{equation}
and $<f>({\bf x})=<f({\bf x},{\bf y})>:=\int_Y f({\bf x},{\bf y})d{\bf y}$ with $Y$ the unit cell.

\noindent Moreover, the effective density reads as
\begin{equation}
\begin{array}{ll}
\rho_0({\bf x})=<\rho>({\bf x}) \; .
\end{array}
\label{toto2}
\end{equation}
\noindent 
One notes that the effective density is thus scalar valued and cannot achieve a rank-2 symmetric density tensor as required by the approximate Willis
medium constituting the cloak, according to (\ref{ddcloakdensor}). Furthermore, it is suggested in \cite{diatta-kadic2016} that at least in some cases, in order for one to perform cloaking with an elastic material defined by  a symmetric elasticity tensor, one might need that the material be also defined by an anisotropic density.
Inspired by earlier work on locally resonant acoustic metamaterials for anisotropic effective density \cite{sashabook,martinsson,torrent1,milton2007,hfh,torrent2,huang,christensen2015}, we therefore add a second step in the homogenization process, that amounts to adding concrete bars in the soil.

\noindent As is well known using spring-mass approximations of locally resonant elastic metamaterials, a dynamic effective (diagonal) density tensor is given by \cite{christensen2015}:
\begin{equation}
\begin{array}{ll}
\rho_{ii}({\bf x})=<\rho>({\bf x})+\Delta_{i} \frac{\omega^2}{\Omega_i^2-\omega^2} \; .
\end{array}
\label{wegener}
\end{equation}
which is a dynamic correction to (\ref{toto2}) with $\Delta_i$ the oscillator strengths of the resonances $\Omega_i$.

\noindent Equipped with (\ref{toto1}) and (\ref{wegener}) for the effective anisotropic elasticity and density tensors, we can now design approximate seismic cloaks with concrete bars adding the required resonant feature to the soil.

\section{Elastic thick plates structured with concrete columns for homogenized Willis medium without rank-3 tensors but anisotropic effective dynamic density}
In this section we investigate the case of seismic cloaks which are a mitigation of cylindrical elastodynamic cloaks as proposed in \cite{brunapl} for in-plane coupled shear and pressure waves (which unfortunately require Cosserat media) and thin plate cloaks \cite{prb2009,farhat2009,farhat2012,stenger2012,andrea1} for Lamb waves. Our case is indeed dedicated to Rayleigh like waves in thick plates. We do not claim that our cloaks work for body waves.

Bearing in mind that many authors have thus far achieved dynamic anisotropic density, see for instance \cite{torrent1,milton2007,torrent2,huang,christensen2015}, symmetrized Willis cloaks are then achievable by classical homogenization approaches (getting rid of rank-3 tensors) with high contrast in material parameters for the artificial anisotropic density.

We would like to explore the propagation of elastodynamic waves in two types of cloaks that are in line with such a high-contrast homogenization, using the finite element method. The two cloaks are dedicated to seismic protection and could be achieved by manufacturing soil using concrete columns. Physical parameters are 
$\lambda_0=87.7*10^6$ Pa, $\mu_0=58.5*10^6$ Pa and density $\rho_0=1800$ kg/m$^3$ for soil and $\lambda_1=18.24*10^9$ Pa, $\mu_1=9.39*10^9$ Pa and density $\rho_1=2300$ kg/m$^3$ for concrete.

\begin{figure}[h!]
\resizebox{84mm}{!}{\includegraphics{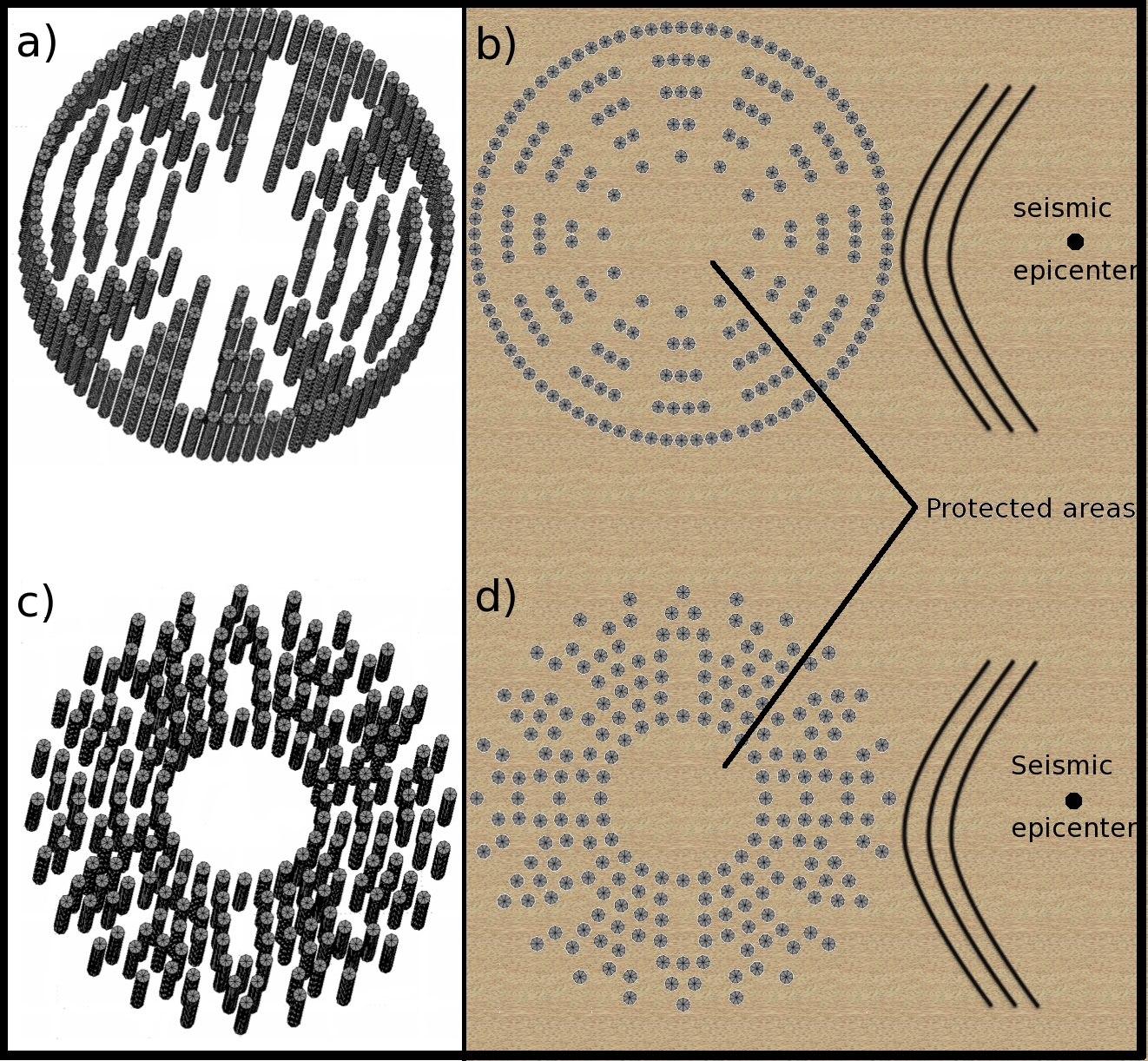}}
\vspace{-3mm}
\caption{Schematic illustration of the structured cloaks of type I (a,b) and type II (c,d) implemented in soil basin. These cloaks are reasonable approximations of ideal
parameters in (\ref{magicmomo}) for cloak of type I (with a decreasing effective density towards the inner boundary of the cloak, as can be seen from the spatial distribution of inclusions in (a)-(b)) and of parameters in (\ref{ddcloaktensor})-(\ref{ddcloakdensor}) for cloak of type II, according to effective equations (\ref{toto1})-(\ref{toto2}), with an increasing anisotropy towards the inner boundary of the cloak (the inner boundary would approximate an infinite anisotropic tensor of elasticity in the limit of densely packed inclusions).}\label{fig1aip}
\end{figure}

In order to prevent eventual reflections on the boundaries, Perfectly Matched Layers (PMLs) are applied to simulate the infinite extent of the elastic medium in the horizontal plane (note that we use so-called adaptative PMLs that work from high-frequencies down to the quasi-static limit, see \cite{diatta-kadic2016}). A point source is placed atop a semi infinite medium representing the earthquake epicenter, which provoke a Rayleigh-like wave (a type of wave that can be harmful and detrimental for buildings). The outside and inside radii of the cloak measure $21.6$ m and $10$ m, respectively. The point source is located at $50$ m from the cloak center. Concrete columns are $2$ m in diameter and penetrate $30$ m in depth. Since we are interested in a frequency range greater than $5$ Hz, the soil depth is set to $50$ m (note that at $5$ Hz Rayleigh waves have a wavelength of $36$ m) and we apply free traction boundary condition to the bottom). We have checked that such a condition gives comparable results to those that are obtained if the bedrock is identified $50$ m beneath soil. Indeed, potential reflection occur in both cases due to impedance mismatch. We deliberately consider such configuration as it seems more realistic than using PMLs that would model an infinitely deep, unstratified, soil. Concrete columns are arranged in a way that the effective density decreases as we move towards the center of the cloak. The designs of structured cloaks are optimized in order to achieve a reasonable balance between approximation of ideal cloak parameters according to effective equations of section IV and what can be implemented in practice with civil engineering technology at hand (i.e. various distributions of concrete columns of various diameter and length have been numerically tested). Indeed, the range of frequencies from $5$ Hz to $10$ Hz corresponds to wavelengths much larger than the columns's diameters ({ subwavelength inclusions underpins} homogenization theory) and small enough compared to the basin depth to investigate Rayleigh waves rather than plate waves. {Since the first cloak design, which we propose is deduced from elements of thin plate theory, we also look at the frequencies from $3$ Hz to $5$ Hz, which correspond to a transition between plate and Rayleigh waves. Indeed, the frequency range of interest for civil engineering is from $1$ to $10$ Hz, and it is interesting to check whether or not our cloaks's designs work beyond their 'legitimate' range of frequencies.}

\subsection{A first cloak design based on thin plate approximation}
Nevertheless, we notice that the distribution of inclusions in Fig. \ref{fig1aip}(a)-(b) is coherent with the formula of the density of an elastic cloak previously reported by \cite{prb2009} in the limit of thin plates (so-called Kirchhoff-Love theory) and reads:

\begin{eqnarray}E_r=\Big(\frac{r-r_1}{r}\Big)^2, ~~  E_\theta=\Big(\frac{r}{r-r_1}\Big)^2, ~~\text{ and } \rho=\alpha^2\Big(\frac{r-r_1}{r}\Big)^2,
\label{magicmomo}
\end{eqnarray}
with $\alpha=\frac{r_2}{r_2-r_1}$.
However, this step doesn't achieve the required effective anisotropy, and should be only valid for plate waves (which are akin to Rayleigh waves in the limit of thick plate not covered by Kirchhoff-Love theory). 

\begin{figure}[h!]
\resizebox{100mm}{!}{\includegraphics{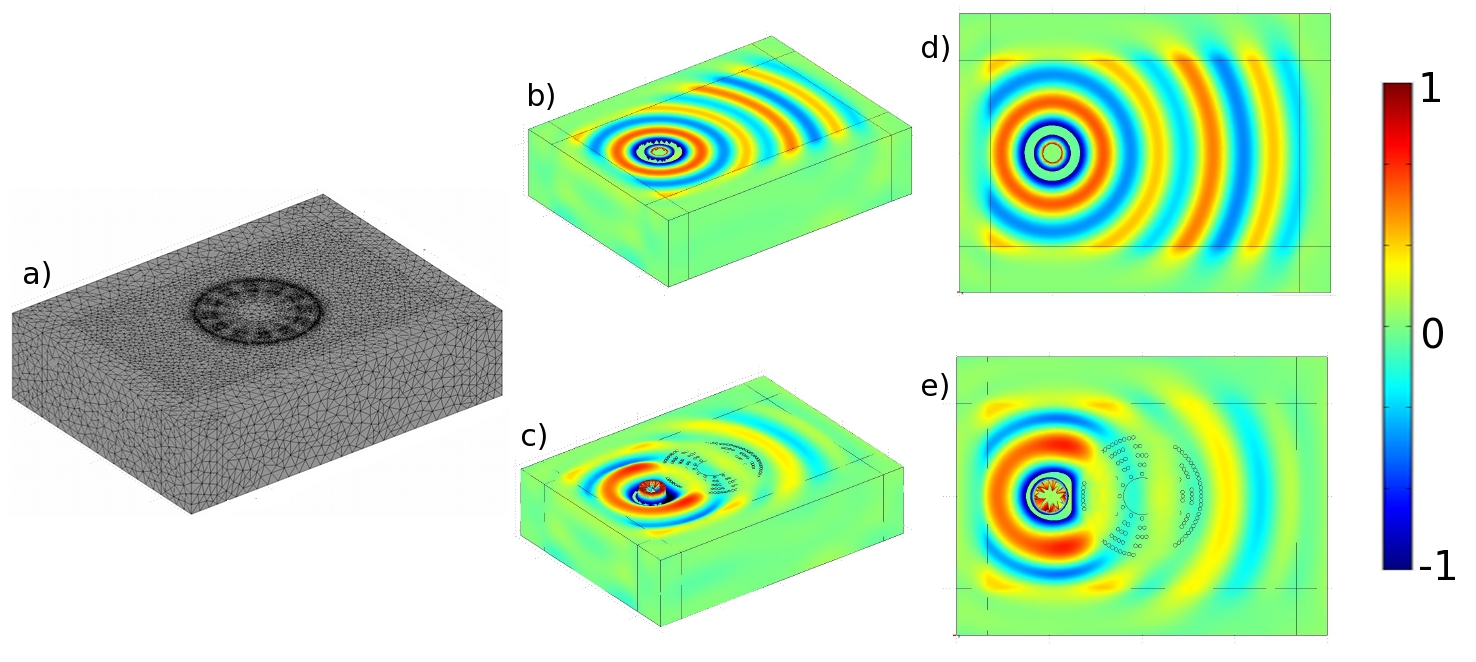}}
\vspace{-3mm}
\caption{Mesh of computational domain (a) with $2,308,656$ degrees of freedom, $562,675$ elements and three-dimensional views (b,c) and top views (d,e) of the out of plane displacement field  (Re($u_z$)) at 5.5 Hz of a Rayleigh-like wave propagating atop a soil without inclusions (b,d) and atop a soil manufactured with cloak of type I (c,e). One notes that the surface wave wavefront is nearly unperturbed by the cloak, but its amplitude is decreased within the center of the cloak.}\label{fig2}
\end{figure}

We depict in Fig. \ref{fig2} the real part of the out of plane displacement field in 3D and top view representations at $5.5$ Hz. It is observed that more than 50\% of the field is suppressed at the center of the cloak (zone of protection) but recovered at its exit. The acceleration phenomenon is also noticed within the cloak area due to the higher effective Young's modulus. However, when we compute the cloak efficiency for protection throughout the frequency range $[3 \hbox{Hz},10 \hbox{Hz}]$,
we find that protection is merely 40\% on average, see Fig. \ref{figlast}(b). This is quite a dramatic reduction of protection compared with the ideal cloak, which has 70\% of protection on average, see Fig. \ref{figlast}(a).

The cloak is efficient between $3$ Hz and $10$ Hz as it transpires in Fig. \ref{figlast}. It is worth noting that the lower limit (3 Hz) is connected to the basin depth (in which case waves transition from Rayeligh to Lamb) and the upper limit (10 Hz) to the diameter of the columns. The smaller the diameter the higher the upper frequency limit. Indeed the phenomenon is constrained by the diffraction that might be caused by the columns as can be observed in Fig. \ref{figall}(a4).

\subsection{A second cloak design based on approximate Willis equations in thick plates}
We now make use of the much more involved transformational (section III) and effective (section IV) models for Willis equations in thick plates. In the configuration of cloak of type II, the inward effective anisotropy increases. This is accompanied by an undesirable reduction of the effective density the outer boundary of the cloak. This can be explained as the following: the concrete columns are arranged by bunch of clusters with one element over the first layer, two elements over the second and so forth up to the sixth layer. The seventh and last layer is isotropic and could be seen as a seismic shield. We depict in Fig. \ref{fig2} the real part of the out of plane displacement field in 3D and top view representations at $5.5$ Hz. For the sake of comparison, the free medium is also represented. We can notice that unlike the first configuration, there is much less reflection at the cloak entrance (due to the smooth mismatch of impedance), a quasi perfect restitution of the field at the exit and a comparable protection inside the cloak (almost 50$\%$ of the energy is suppressed on average in the frequency interval $[5 \hbox{Hz},10 \hbox{Hz}]$ as can be seen in Fig. \ref{figlast}(c)). However, it is noticed in Fig. \ref{figlast}(a), that some resonances occur in the ideal cloak, and similar effects occur with cloak of type II (notably around $8.5$ Hz) in Fig. \ref{figlast}(c).

\begin{figure}[h!]
\resizebox{100mm}{!}{\includegraphics{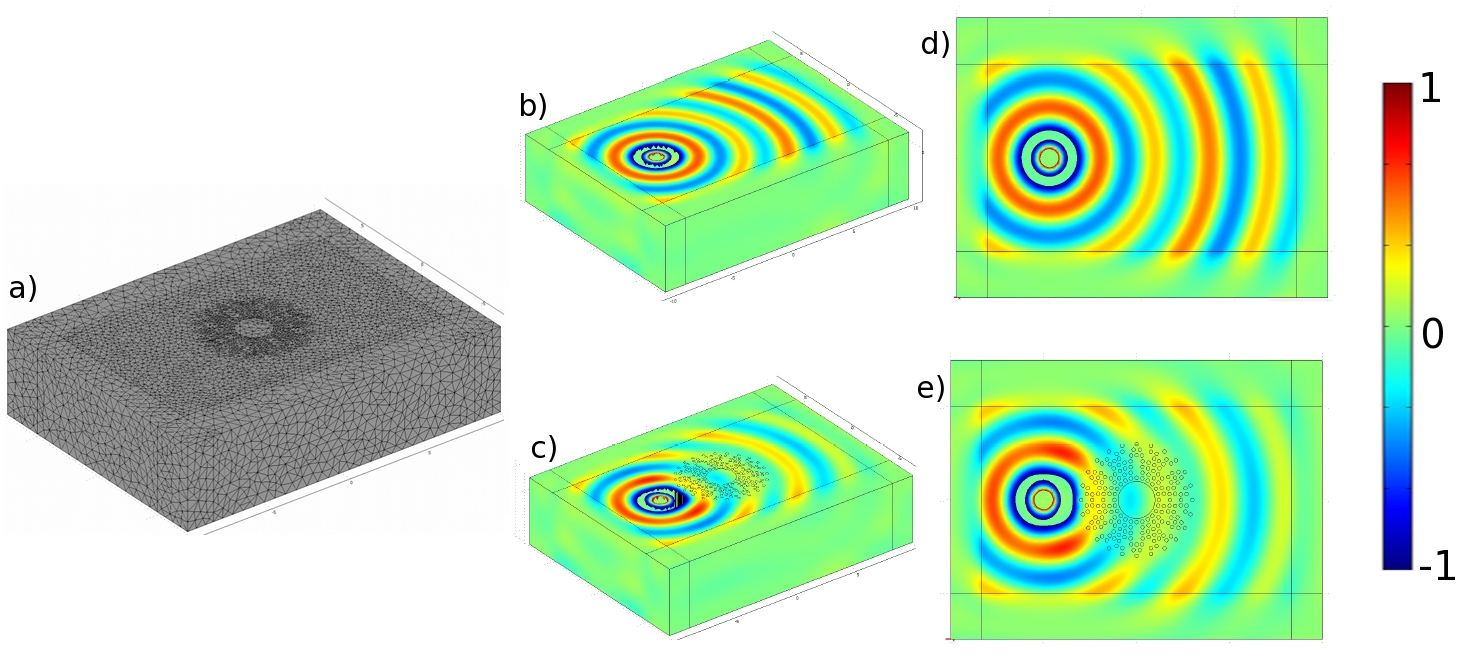}}
\vspace{-3mm}
\caption{Same as Fig. \ref{fig2} with soil structured by a cloak of type II. Note the improved protection in the center of the cloak and the improved restitution of the field on the exit of the cloak compared with cloak of type I in Fig. \ref{fig2}.}\label{fig5}
\end{figure}

\begin{figure}[h!]\label{fig3}
\resizebox{140mm}{!}{\includegraphics{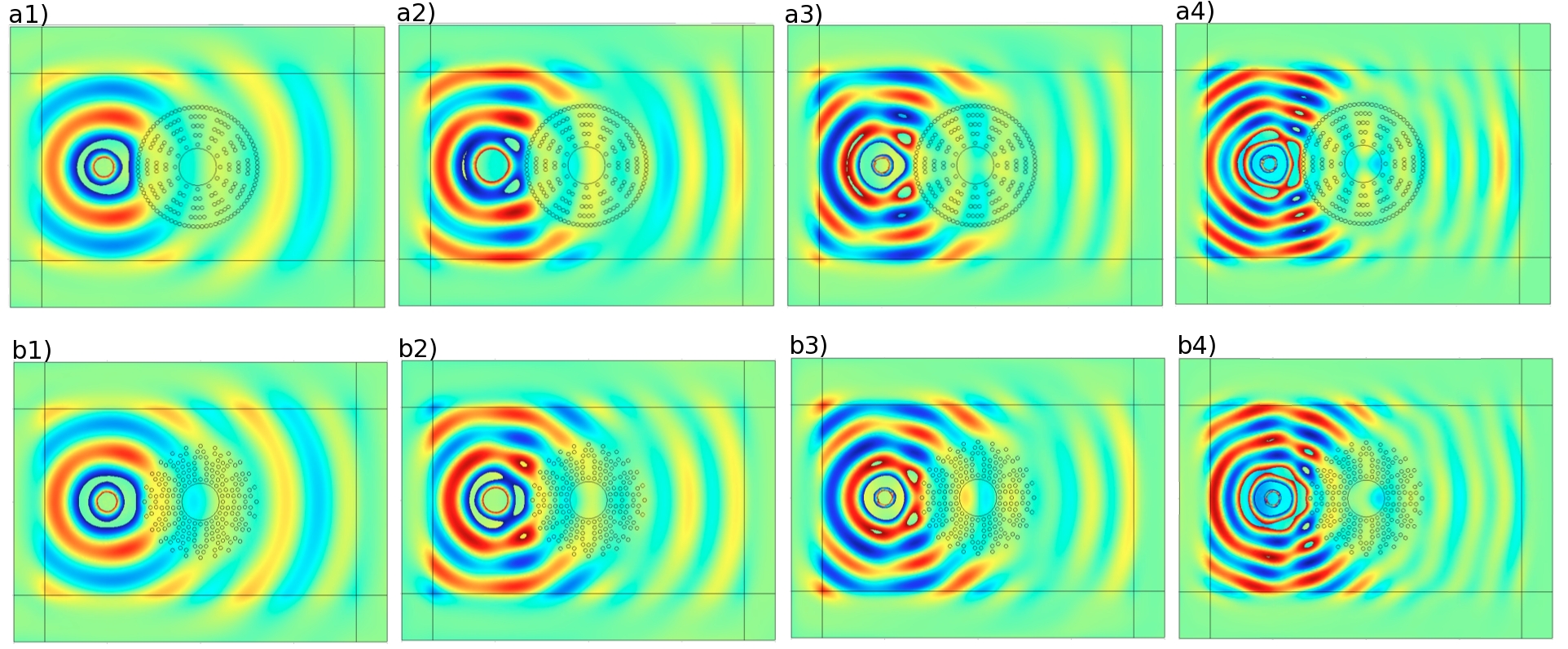}}
\vspace{-3mm}
\caption{Top views of out of plane displacement field (Re($u_z$)) at $5.5$ Hz (a1,b1), $7$ Hz (a2,b2), $8$ Hz (a3,b3) and $10$ Hz (a4,b4). One notes that the wave wavefront is nearly unperturbed by the cloak of type II, and its amplitude is lower within the center of the cloak II than in the center of cloak I.}
\label{figall}
\end{figure}

\subsection{First conclusions on cloaks of type I and II}
As can be seen in Fig. \ref{figall}, cloak of type II performs clearly better than cloak of type I for both protection and invisibility. This can be attributed to the fact that the latter has been designed using transformed equations \cite{farhat2012}, and homogenization techniques \cite{farhat2009} in the context of thin plate theory. Kirchhoff-Love theory is an approach well suited for control of Lamb waves as experimentally validated in \cite{stenger2012}. However, such a limit theory is not legitimate in our case of thick plates, at least from
a mathematical standpoint (the Rayleigh-like wave wavelength should be at least ten times smaller than the plate thickness to satisfy the hypothesis of thin plate, which is clearly not the case in our study). Interestingly, both cloaks of type I and II display resonances similarly to ideal cloak, see Fig. \ref{figlast} (with a better match between ideal cloak and cloak of type II). We believe such unwanted resonances require a specific analysis of its own, as addressed in the next section. It is interesting to note that the level of protection displayed by our three cloaks (both ideal cloak and cloaks of types I and II) is lower than what is achieved through shielding effects via low frequency stop bands in \cite{miniaci16}. Although we consider the same frequency range as in  \cite{miniaci16}, in our case we detour Rayleigh waves around the protected area, and there is little backscattered wave, so the protection mechanism is radically different.

\section{Estimate of local resonances inside the seismic cloak}
\begin{figure}[h!]
\resizebox{84mm}{!}{\includegraphics{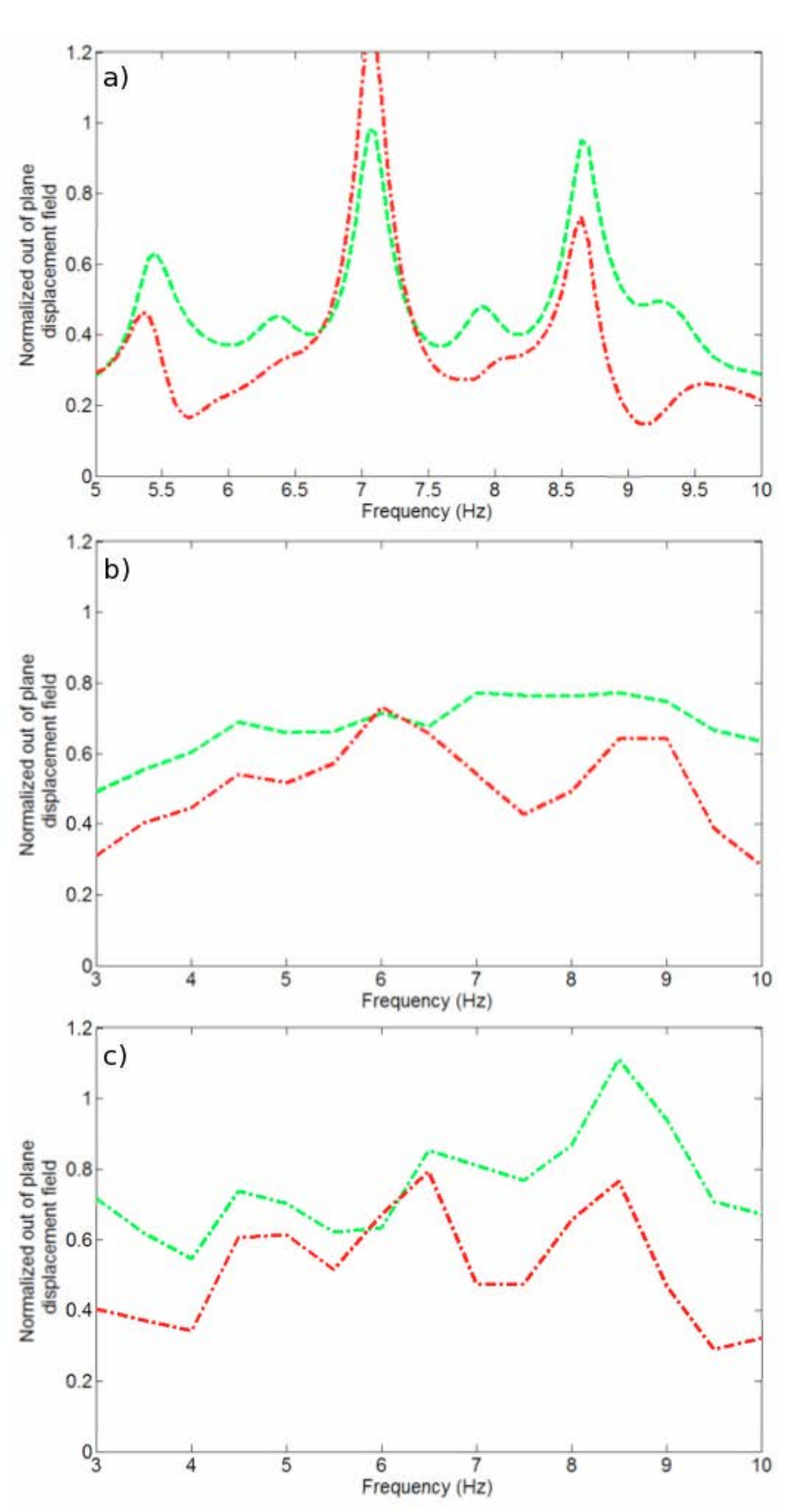}}
\vspace{-3mm}
\caption{Wave protection performances of the ideal cloaks (a) and approximate cloaks of types I (b) and II (c). Red (resp. green) curves correspond to integral of out-of-plane displacement computed over the surface (resp. over the volume) of the invisibility region (i.e. center of cloak) normalized by same in free space. Note that Rayleigh wave wavelength is $60$ m at $3$ Hz and $36$ m at $5$ Hz, whereas plate thickness is $50$ m. Therefore, the frequency range $[3 \hbox{Hz},5 \hbox{Hz}]$ corresponds to
a transition between plate and Rayleigh-like waves. Note also the strong similarities between resonances in (a) and (c); Indeed, cloak of type II in (c) is a better approximation of ideal cloak in (a), than cloak of type I in (b), the latter being designed using thin plate theory. Achieved protection is $70\%$ on average in (a), $40\%$ in (b) and $50\%$ in (c).}
\label{figlast}
\end{figure}

Following \cite{diatta2016}, where the case of Cosserat spherical cloaks is dealt with, in this section we investigate
the origin of the resonances in the ideal cylindrical cloak {by simplifying the governing Navier equations into a simple spring-mass model approximation leading to a transcendental equation}. To do so, we assume that the plate thickness goes to infinity, in which case one can decouple the Navier equations as a vector problem for in-plane shear and pressure waves and a scalar out-of-plane problem for the vertical displacement $u_z$. By doing so, we do not exactly address the problem of the Rayleigh waves, but this will nonetheless give us a reasonable estimate of the resonances, as we shall see. Because of the symmetry of the cloak, one can assume the out-of-plane displacement has a form $u_z=u_z(r')$, solution of the scalar Helmholtz equation
\begin{equation}
\begin{array}{lll}
&\displaystyle{\frac{1}{r'}\frac{\partial}{\partial r'}\left(r'\mu_r\frac{\partial}{\partial r'}u_z\right)+\frac{\mu_\theta}{{r'}^2}\frac{\partial^2}{\partial \theta^2}u_z}
+\displaystyle{\omega^2\rho_z u_z=0}
\end{array}
\label{bleu}
\end{equation}
where $\mu_r(r')=\mu_0{(r'-r_1)}/r'$, $\mu_\theta(r')=\mu_0 r'/(r'-r_1)$ and $\rho_z(r')=\rho_0({(r'-r_1)}/r') (r_2^2/{(r_2-r_1)}^2)$.
Actually,  (\ref{bleu}) can be written as a Bessel equation so it has a general solution, which is expressed in terms of cylindrical Bessel functions of first and second kinds (already derived in a different context in \cite{guenneau2010}):
\begin{equation}
u_z(r')=AJ_0\left(\frac{(r'-r_1)\omega\rho_0 {r}_2}{\mu_0{(r_2-r_1)}}\right)+BY_0\left(\frac{(r'-r_1)\omega\rho_0 {r}_2}{\mu_0{(r_2-r_1)}}\right) \; .
\end{equation}
Then, assuming that $du_z(r_1)/dr=0$ and $u_z(r_2)=0$, we find that the eigenfrequency should
satisfy the following equation
\begin{equation}\label{resonances}
Y_0(\omega r_2 \sqrt{\mu_0/\rho_0})/J_0(\omega r_2 \sqrt{\mu_0/\rho_0})=0 \; ,
\end{equation}
which gives the frequency estimate $f=\omega/(2\pi)\sim 6$ Hz since the first zero of $Y_0(x)$ is $3.8317$. This frequency estimate is in good agreement with the moderate elastic field enhancement shown in Fig. \ref{figlast} around $5.3$ Hz. The frequency estimate for the second resonance is  $f=\omega/(2\pi)\sim 11$ Hz since the second zero of $Y_0(x)$ is $7.0156$ and this is also in qualitatively good agreement with the second peak in Fig. \ref{figlast} around $7.2$ Hz. In fact, the predicted frequencies are overestimated since they are associated with out-of-plane shear waves, that propagate faster than Rayleigh waves.

\section{Concluding remarks}
In this article, we have conclusively shown that transformed Navier equations of the Willis type with a symmetric transformed elasticity tensor, allows for approximate cloak's designs getting rid of the rank-3 transformed tensors unveiled in \cite{milton2006}. A further approximation consists in structuring soil with buried concrete pillars judiciously placed (note that this is the opposite case to what is proposed in \cite{andrea}, wherein soil was structured with softer columns). This is done and finite element computations confirm the cloaking efficiency (invisibility and protection) for two types of large scale seismic cloaks within a soft elastic plate (with soil parameters) $50$ m deep reinforced with judiciously placed columns of concrete $20$ m long and $2$ m in diameter. Indeed, the two designs which we propose allow one to considerably reduce the elastic field vibrations in the center of the cloak with virtually no disturbance of the wave field outside the cloak. The protection could be further improved using the concept of mixed cloak \cite{diatta2016}, which amounts to adding a PML layer at the inner boundary of the cloak, in essence some absorptive anisotropic medium. Of course, alternative routes exist to seismic wave protection, and we would like to mention very promising designs of seismic metamaterials in the tracks of M\'enard's 2012 field test experiment in Grenoble \cite{prl2014}, notably so-called seismic waveguides \cite{kim2012}, but also large scale isochronous mechanical oscillators \cite{finochhio14}, inertial resonators \cite{krodel15,achaoui2015,eml2016} and auxetic-like foundations allowing for low frequency stop bands.
Actually, in \cite{miniaci16}, numerical analysis of both surface and guided waves has been performed in miscellaneous large scale phononic crystals and metamaterials, with soil dissipation effects in viscoelastic media. Protection has been shown for frequency ranges associated with low frequency stop bands below $10$ Hz. Such design of seismic shields could be combined with seismic cloaks to offer a wave protection for both surface and bulk waves which does not have adverse effects for the surrounding environment. Speaking of which the idea of forest of trees for seismic wave mitigation remains the most environmental friendly proposal \cite{colombi}, although it has been observed thus far that wave protection is achieved beyond the $1$ to $10$ Hz frequency range of interest for civil engineers. 

\begin{figure}[h!]
\resizebox{84mm}{!}{\includegraphics{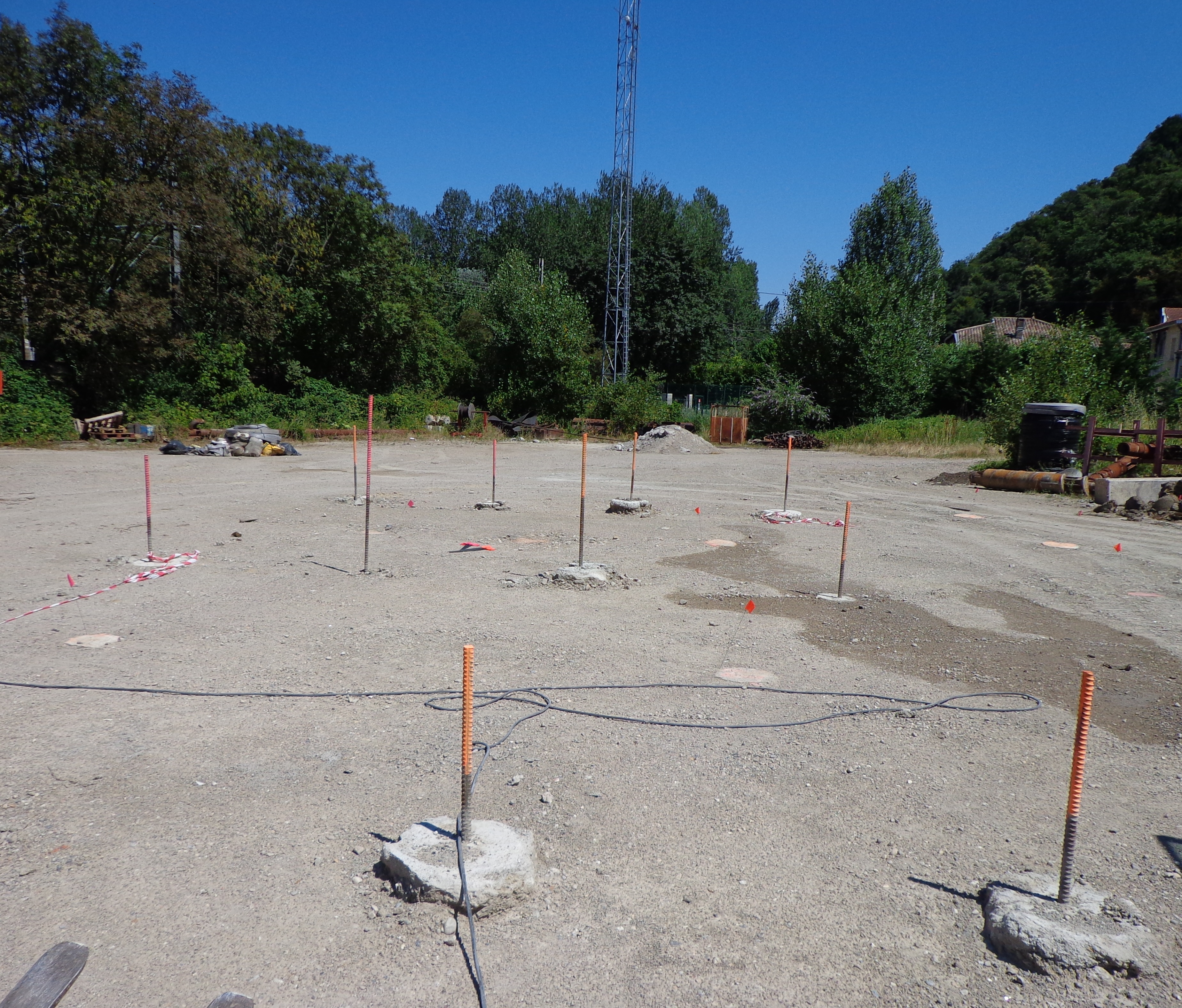}}
\vspace{-3mm}
\caption{Photo of a soil reinforced with columns of concrete $0.4$m in diameter and $15$m in depth (Courtesy of M\'enard).}
\label{Menard}
\end{figure}

We would like to add that our results could be translated into geophysics upon use of time domain computations. It would be also important to study the effect of viscoelasticity on the elastic wave propagation (in a way similar to what was done in \cite{miniaci16}), which can be done with comsol multiphysics. Finally, some implementation of the seismic cloak with subsequent field test is in progress with the M\'enard company.
We show in Figure \ref{Menard} a typical field test experiment by the M\'enard company that could serve as a basis for the validation of our concept of seismic cloak, although this requires scaling down the diameter of inclusions of our study from $2$ m to $0.4$ m, as well as taking into account other constraints on minimum distance between inclusions, field topography etc. New designs of seismic cloaks are therefore underway to find a good compromise between cloak efficiency and feasibility in practice.

\section*{Acknowledgments}
The authors would like to thank the anonymous reviewer for insightful remarks. 
A.D., Y.A. and S.G. acknowledge European funding through ERC Starting Grant ANAMORPHISM.

\end{document}